\documentclass[prl,twocolumn,superscriptaddress,amsmath,amssymb]{revtex4-1}

\usepackage{braket,mathtools}
\usepackage{upgreek}
\usepackage[unicode=true]{hyperref}
\usepackage{subfigure,bbm,times}
\usepackage[T1]{fontenc}

\renewcommand{\tilde}{~}

\newcommand{\trace}[1]{\text{Tr}\left[#1\right]}

\newcommand{\kluru}{\ket{L\uparrow,R\uparrow}}
\newcommand{\klurd}{\ket{L\uparrow,R\downarrow}}
\newcommand{\kldru}{\ket{L\downarrow,R\uparrow}}
\newcommand{\kldrd}{\ket{L\downarrow,R\downarrow}}

\newcommand{\klulu}{\ket{L\uparrow,L\uparrow}}
\newcommand{\kluld}{\ket{L\uparrow,L\downarrow}}
\newcommand{\kldld}{\ket{L\downarrow,L\downarrow}}
\newcommand{\kruru}{\ket{R\uparrow,R\uparrow}}
\newcommand{\krurd}{\ket{R\uparrow,R\downarrow}}
\newcommand{\krdrd}{\ket{R\downarrow,R\downarrow}}

\newcommand{\rlr}{\rho_{\scriptscriptstyle{\text{LR}}}}

\newcommand{\rbs}{\rho_{\scriptscriptstyle\text{BS}}}

\newcommand{\rno}{\rho_{\scriptscriptstyle\text{NO}}}

\newcommand{\plr}{p_{\scriptscriptstyle\text{LR}}}
\newcommand{\pno}{p_{\scriptscriptstyle\text{NO}}}

\newcommand{\kop}[1]{\ket{1_+}_{\scriptscriptstyle\textrm{#1}}}
\newcommand{\kom}[1]{\ket{1_-}_{\scriptscriptstyle\textrm{#1}}}
\newcommand{\ktp}[1]{\ket{2_+}_{\scriptscriptstyle\textrm{#1}}}
\newcommand{\ktm}[1]{\ket{2_-}_{\scriptscriptstyle\textrm{#1}}}
\newcommand{\kup}[1]{\ket{U_+}_{\scriptscriptstyle\textrm{#1}}}
\newcommand{\kum}[1]{\ket{U_-}_{\scriptscriptstyle\textrm{#1}}}
\newcommand{\kdp}[1]{\ket{D_+}_{\scriptscriptstyle\textrm{#1}}}
\newcommand{\kdm}[1]{\ket{D_-}_{\scriptscriptstyle\textrm{#1}}}

\newcommand{\bum}[1]{\bra{U_-}_{\scriptscriptstyle\textrm{#1}}}

\newcommand{\bdm}[1]{\bra{D_-}_{\scriptscriptstyle\textrm{#1}}}
\newcommand{\bop}[1]{\bra{1_+}_{\scriptscriptstyle\textrm{#1}}}
\newcommand{\bom}[1]{\bra{1_-}_{\scriptscriptstyle\textrm{#1}}}
\newcommand{\btp}[1]{\bra{2_+}_{\scriptscriptstyle\textrm{#1}}}
\newcommand{\btm}[1]{\bra{2_-}_{\scriptscriptstyle\textrm{#1}}}

\newcommand{\proj}[1]{\Pi_{\scriptscriptstyle\text{#1}}}
\newcommand{\base}[1]{\mathcal{B}_{\scriptscriptstyle\text{#1}}}

\newcommand{\so}{\mathcal{S}_1}
\newcommand{\st}{\mathcal{S}_2}

\newcommand{\rdep}{\rho_{\text{dep}}}

\begin{document}
    
    \title{Asymptotically-deterministic robust preparation of maximally entangled bosonic states}

    \author{Matteo Piccolini}
	\email{matteo.piccolini@unipa.it}
	\affiliation{Dipartimento di Ingegneria, Universit\`{a} di Palermo, Viale delle Scienze, 90128 Palermo, Italy}
	
    \author{Vittorio Giovannetti}
        \affiliation{NEST, Scuola Normale Superiore and Istituto Nanoscienze-CNR, I-56126 Pisa, Italy}
	
    \author{Rosario Lo Franco}
	\email{rosario.lofranco@unipa.it}
	\affiliation{Dipartimento di Ingegneria, Universit\`{a} di Palermo, Viale delle Scienze, 90128 Palermo, Italy}

    \begin{abstract}
        We introduce a theoretical scheme to prepare a pure Bell singlet state of two bosonic qubits, in a way that is robust under the action of arbitrary local noise. Focusing on a photonic platform, the proposed procedure employs passive optical devices and a polarization-insensitive, non-absorbing, parity check detector in an iterative process which achieves determinism asymptotically with the number of repetitions.
        Distributing the photons over two distinct spatial modes, we further show that the elements of the related basis composed of maximally entangled states can be divided in two groups according to an equivalence based on passive optical transformations. We demonstrate that the parity check detector can be used to connect the two sets of states. We thus conclude that the proposed protocol can be employed to prepare any pure state of two bosons which are maximally entangled in either the internal degree of freedom (Bell states) or the spatial mode (NOON states).
    \end{abstract}

    \maketitle

    \textit{Introduction.}
        Entanglement, the most exotic property of quantum mechanics, is at the heart of the enhancement provided by quantum protocols in many different fields of application\tilde\cite{horodecki}, ranging from metrology and parameter estimation\tilde\cite{giovannetti2011advances,paris2009quantum}, to computation\tilde\cite{ladd2010quantum}, communication, and cryptography\tilde\cite{pirandola2020advances}.
        The ability to prepare entangled states with high reliability is thus crucial for the practical development of quantum technologies. Nonetheless, realistic preparations of entangled states are known to be hindered by the ubiquitous interaction with the surrounding environment, whose noisy action is detrimental for the quantum correlations within the system\tilde\cite{horodecki,breuer2002theory,aolita_2015}. For this reason, many different techniques to circumvent the problem have been proposed over time\tilde\cite{preskill_1998,knill2005quantum, shor_1995,steane_1996,mazzola_2009,bellomo_2008,lo_franco_2013,aolita_2015,
        Xu_2010,bylicka_2014,man_2015,tan_2010,tong_2010,
        breuer_2016_colloquium,man_2015_pra,Bennett1996,
        horodecki1997inseparable,horodecki1998mixed,horodecki2001distillation,
        kwiat_2001,dong_2008,zanardi1997noiseless,lidar1998decoherence,
        Viola1998,viola2005random,darrigo_2014_aop,franco2014preserving,
        orieux_2015,facchi_2004,lo_franco_2012_pra,xu_2013,damodarakurup_2009,
        cuevas_2017,Mortezapour_2017,Mortezapour_2018}.

        In this Letter, we first propose a protocol to distil a pure, maximally entangled Bell singlet state of two bosons from a completely depolarized one. We focus on a photonic implementation. The local action of depolarizing channels, which can be efficiently induced by randomized local polarization rotators, transforms any arbitrary state of spatially distinguishable photons into a maximally mixed one. Thus, the proposed procedure can be applied to \textit{any} arbitrary \textit{initial state} of two spatially distinguishable photons, regardless of local noises affecting them before the depolarization.
        Our scheme employs  passive optical (PO) devices and a polarization-insensitive, non-absorbing, parity check detector. The latter is an highly nonlinear transformation which performs a quantum nondemolition (QND) measurement capable to discriminate between states with even/odd number of photons. More precisely we only require the detector to distinguish between the cases where in a given location we have a single photon (corresponding to the successful generation of the Bell singlet), and those in which the total number of photons is either zero or equal to two (corresponding to a failure). The nondemolition character of the measurement ensures that in case of failure the whole protocol can be repeated by depolarizing the system once again, resetting it to the maximally mixed state.
     By doing so, the preparation of the Bell singlet is achieved with a probability scaling to 1 exponentially with the number of repetitions, thus being \textit{asymptotically-deterministic}. 

   \begin{figure}[b!]
            \includegraphics[width=0.95\columnwidth]{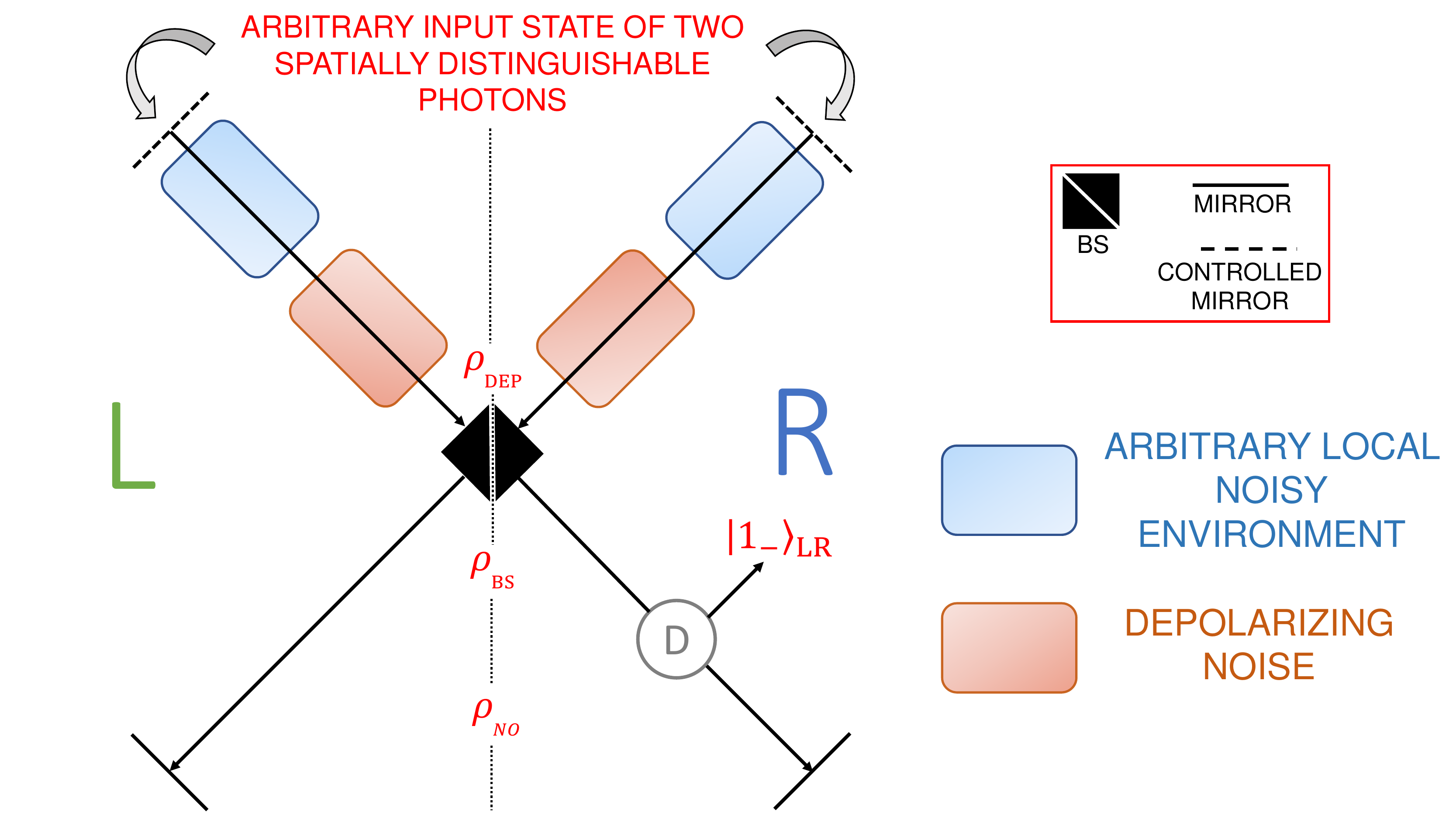}
            \centering
                \caption{\textbf{Schematic representation of the setup.} The D element represents a polarization-insensitive, non-absorbing parity check detector. The depolarization noises in the red area (just before the BS) are assumed to be externally activated, while the noise sources in the blue area are environmentally induced.}
        \label{procedure}
        \end{figure}

        Differently from other entanglement distillation protocols\tilde\cite{Bennett1996,horodecki1997inseparable,horodecki1998mixed,horodecki2001distillation} allowing only for local operations and classical communication (LOCC), our scheme makes explicit use of the interference effects due to particle indistinguishability when non-locality is generated by a beam splitter (BS). In this sense, the proposed procedure extends the results obtained in Refs.\tilde\cite{indistentanglprotection,Piccolini_2021_entropy,piccolini2021opensys}, where the authors employed a technique based on \textit{spatially localized operations and classical communication} (sLOCC)\tilde\cite{slocc,piccolini2022philtrans,experimentalslocc,Barros:20,sun2022activation,wang2022remote,wang2022proof} to achieve a probabilistic distillation of a Bell singlet state from a singlet subjected to the action of local noisy environments.
        
        Finally, we introduce an equivalence between bosonic bipartite states based on PO transformations. We consider an orthonormal basis of the bipartite Hilbert space composed of only maximally entangled states, and show that it can be divided in two sets of PO equivalent elements. We demonstrate that the two sets can be connected by means of the polarization-insensitive, non-absorbing, parity check detector previously discussed. As the Bell singlet state belongs to one of the two sets, we thus conclude that the proposed procedure allows for the preparation of \textit{any} arbitrary \textit{maximally entangled, pure bipartite state}. This comes with a trade-off in the difficult realization of the exotic detector required, whose crucial role for quantum information protocols emerges by the relevance of the reported results themselves.

        Identical particles are treated via the \textit{no-label approach}\tilde\cite{nolabelappr,compagno2018dealing,slocc}, a mathematical framework which allows to overcome some of the main issues affecting the standard label-based formalism\tilde\tilde\cite{tichy,ghirardi}. Also, it allows to write multiparticle states without having to explicitly symmetrize/antysimmetrize them as ruled by the symmetrization postulate\tilde\cite{nolabelappr,compagno2018dealing,slocc}, thus simplifying the notation.
        
    \textit{Notation.}
        The Hilbert space of two bosonic qubits distributed over two distinct spatial regions L and R is 10-dimensional. We consider a basis $\mathcal{B}=\base{LR}\cup\base{NO}$ of maximally entangled states, where

        \begin{equation}
        \label{basis}
            \begin{aligned}
                \base{LR}
                &:=\Big\{
                    \ket{1_\pm}_{\scriptscriptstyle\textrm{LR}},\,\ket{2_\pm}_{\scriptscriptstyle\textrm{LR}}
                \Big\},
                \\
                \base{NO}
                &:=\Big\{
                    \ket{1_\pm}_{\scriptscriptstyle\textrm{NO}},\,\ket{U_\pm}_{\scriptscriptstyle\textrm{NO}},\,
                    \ket{D_\pm}_{\scriptscriptstyle\textrm{NO}}
                \Big\},
            \end{aligned}
        \end{equation}
        and
        \begin{equation}
            \begin{gathered}
                \ket{1_\pm}_{\scriptscriptstyle\textrm{LR}}
                :=\frac{1}{\sqrt{2}}\Big(\klurd\pm\kldru\Big),\\
                \ket{2_\pm}_{\scriptscriptstyle\textrm{LR}}
                :=\frac{1}{\sqrt{2}}\Big(\kluru\pm\kldrd\Big),\\
                \ket{1_\pm}_{\scriptscriptstyle\textrm{NO}}
                :=\frac{1}{\sqrt{2}}\Big(\kluld\pm\krurd\Big),
                \\
                \ket{U_\pm}_{\scriptscriptstyle\textrm{NO}}
                :=\frac{1}{2}\Big(\klulu\pm\kruru\Big),
                \\
                \ket{D_\pm}_{\scriptscriptstyle\textrm{NO}}
                :=\frac{1}{2}\Big(\kldld\pm\krdrd\Big).
            \end{gathered}
        \end{equation}
        
        Notice that the elements of basis $\base{LR}$ are Bell states entangled in the internal degree of freedom $\ket{\uparrow},\,\ket{\downarrow}$, which, for the photonic implementation considered in the
        following paragraphs, can be identified with the polarization; instead, the basis $\base{NO}$ is composed of NOON states entangled in the spatial degree of freedom.

    \textit{Procedure.}
        The proposed scheme is depicted in Fig.\tilde\ref{procedure}.
        Let us take an arbitrary state of two photons localized in two distinct spatial modes L and R. If each photon is locally subjected to a depolarizing channel that induces a complete randomization of its polarization degree of freedom, such a state will be mapped into a maximally mixed configuration which can be expressed as a uniform mixture of the elements of the basis $\base{LR}$ introduced above, i.e., $\rdep:=\frac{1}{4}\,\proj{LR}$, where $\proj{LR}:=\sum_{\ket{v}\in\base{LR}}\ket{v}\bra{v}$ is the projector onto the subspace spanned by the elements of the basis $\base{LR}$.
        We now let the two photons impinge on the two input ports of a balanced beam splitter (BS), which mixes the L and R regions inducing, at the level of single particle states, the mappings $\ket{L}\longrightarrow(\ket{L}+\ket{R})/\sqrt{2}$ and $\ket{R}\longrightarrow(\ket{L}-\ket{R})/\sqrt{2}$.
        Applied to the elements of the set $\base{LR}$, this achieves the transformations
        \begin{equation}
        \label{bsaction}
            \left\{
            \begin{aligned}
                &\kom{LR}\longleftrightarrow-\kom{LR},\\
                &\kop{LR}\longleftrightarrow\kom{NO},\\
                &\ktm{LR}\longleftrightarrow(\kum{NO}-\kdm{NO})/{\sqrt{2}},\\
                &\ktp{LR}\longleftrightarrow(\kum{NO}+\kdm{NO})/{\sqrt{2}}.
            \end{aligned}
            \right.
        \end{equation}
        As a result, the state $\rdep$ introduced previously is mapped into 
        \begin{equation}
        \label{rbs}
            \rbs=\frac{1}{4}\kom{LR}\bom{LR}+\frac{3}{4}\rno,
        \end{equation}
        where
        \begin{equation}
        \label{rno}
            \rno:=
            \frac{1}{3}
            \Big(\kom{NO}\bom{NO}+\kum{NO}\bum{NO}+\kdm{NO}\bdm{NO}\Big).
        \end{equation}
        We highlight that $\rbs$ in Eq.\tilde\eqref{rbs} is a classical mixture of the Bell singlet state $\kom{LR}$ and of NOON states. Crucially, the former is characterized by an odd number of photons in each spatial mode, while an even number (0 or 2) characterizes the latter. This fact can be exploited to distil the singlet as follows. We employ a polarization-insensitive, non-absorbing, parity check detector D. By monitoring one of the two spatial modes, such a detector is capable to distinguish whether it contains an odd or an even number of photons. In the first case, $\rbs$ is projected onto the subspace spanned by the Bell states composing $\base{LR}$ via the projection operator $\proj{LR}$ previously introduced,
        giving the desired singlet $\kom{LR}$. In this case, occurring with probability $\plr=\trace{\proj{LR}\rbs}=1/4$, we collect the state and conclude the process. If D registers an even number of photons, instead, $\rbs$ is projected onto the subspace spanned by the NOON states in basis $\base{NO}$ via the projection operator $\proj{NO}:=\sum_{\ket{k}\in\base{NO}}\ket{k}\bra{k}.$
        This scenario, which occurs with probability $\pno=\trace{\proj{NO}\rbs}=3/4$, leaves the system in the state $\rno$ of Eq.\tilde\eqref{rno}. In this case, we act on the system with another beam splitting operation, obtaining the state
        \begin{equation}
        \label{xino}
            \xi_{\scriptscriptstyle\text{LR}}:=
            \frac{1}{3}
            \Big(\kop{LR}\bop{LR}+\ktp{LR}\btp{LR}+\ktm{LR}\btm{LR}\Big).
        \end{equation} The two photons are now subjected to local depolarizing channels once again, resetting the system to the completely depolarized state $\rdep$ we started with. The process can thus be repeated a second time without having to inject new photons in the setup, leading to the generation of a Bell singlet state with total probability $\plr^{(2)}=1/4+(3/4)(1/4)$. Proceeding this way, the $j$-th iteration returns $\kom{LR}$  with probability $\plr^{(j)}=\sum_{n=1}^{j}(1/4)(3/4)^{n-1}$, which converges exponentially to 1 for $j\to\infty$.

    \textit{Amplitude damping-based implementation.}
        Here we propose an alternative implementation of our scheme which adopts two local amplitude damping channels instead of the depolarizing ones. In this case, the noisy environments map two spatially separated qubits into the pure ground state $\kldrd$. Placing a polarization rotator (PR) (see below) on the spatial mode L (to fix a framework), we get $\klurd=(\kop{LR}+\kom{LR})/\sqrt{2}$. From Eq.\tilde\eqref{bsaction}, we notice that the BS transforms this state into $(\kom{NO}-\kom{LR})/\sqrt{2}$. The detector D can now be employed to distill a Bell singlet state with probability $\plr=1/2$. When the system is found in state $\kom{NO}$, instead, the process is repeated analogously to the case where depolarizing channels are employed. At the $j$-th iteration, the singlet is distilled with probability $\plr^{(j)}=\sum_{n=1}^{j} 1/2^{n}$, which again converges to 1 exponentially when $j\to\infty$.

    \textit{Passive optical equivalence.}
        We introduce \textit{PO operations} as the set of transformations which can be obtained by a proper sequence of BSs, polarization BSs (PBSs), polarization-dependent or -independent phase shifters (PDPSs/PIPSs), and local polarization rotators (PRs). We further define two states to be \textit{PO equivalent} if they can be obtained one from the other by means of PO operations.

\begin{figure}[t!]
            \includegraphics[width=0.95\columnwidth]{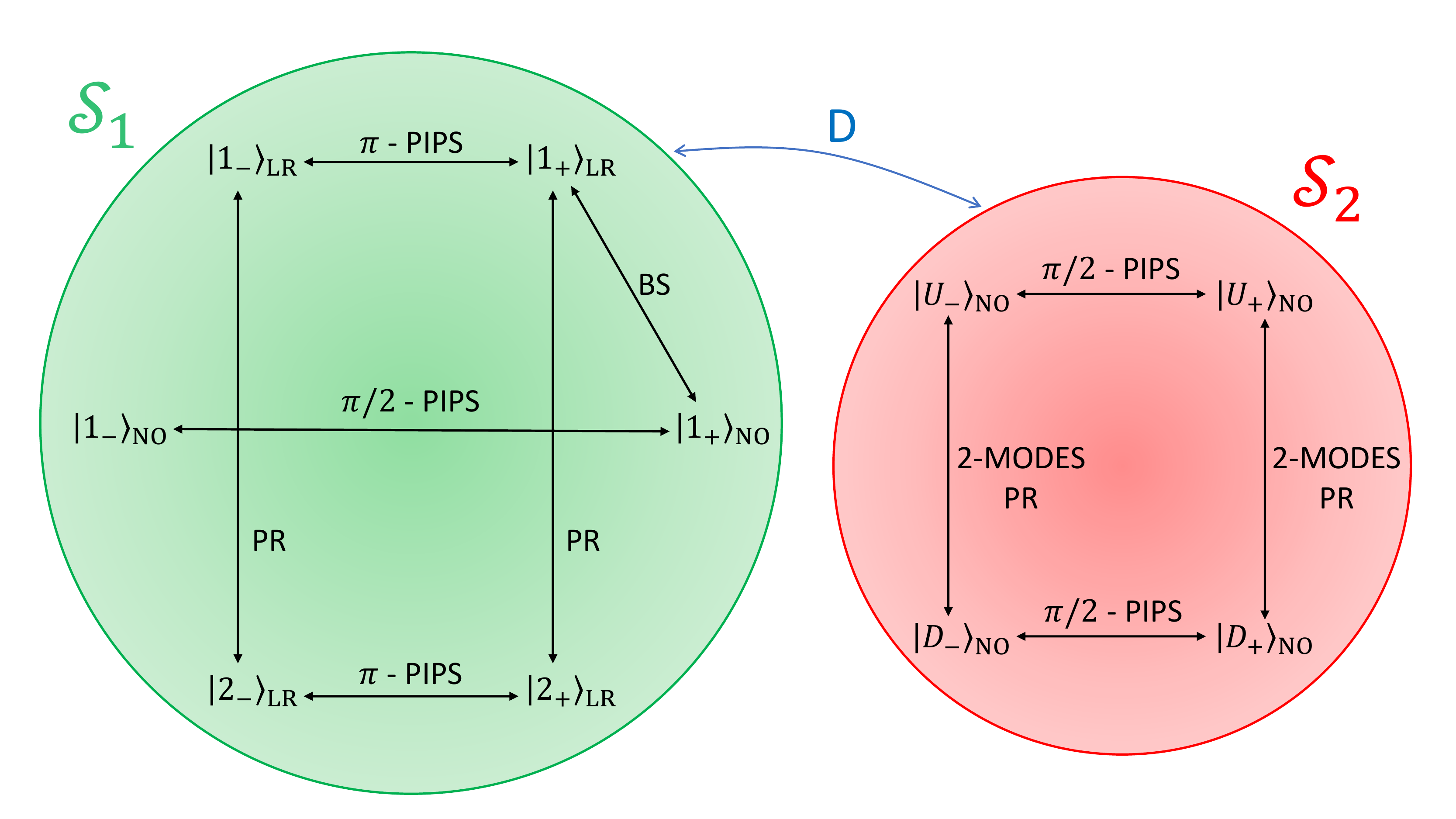}
            \centering
                \caption{\textbf{Structure of passive optical equivalent maximally entangled states of two photons.} The figure shows two sets of PO equivalent maximally entangled states of two bosonic qubits distributed over two spatial modes. Examples of PO transformations connecting them are reported for each set. All the depicted PO transformations are assumed to occur on a single arbitrary spatial mode, except when 2-modes is stated. $\theta$-PIPS are polarization independent phase shifters inducing a phase $\theta$ on the spatial mode they are set on, PRs are $90^\circ$ polarization rotators, and BSs are beam splitters. The two sets are linked by a polarization-insensitive, non-absorbing, parity check detector D (see main text).}
        \label{poscheme}
        \end{figure}

        PO equivalence allows to divide basis $\mathcal{B}$ in Eq.\tilde\eqref{basis} in two sets of equivalent states:
        \begin{equation}
        \label{equisets}
            \begin{gathered}
                \so:=\Big\{\ket{1_\pm}_{\scriptscriptstyle\textrm{LR}},\,\ket{2_\pm}_{\scriptscriptstyle\textrm{LR}},\ket{1_\pm}_{\scriptscriptstyle\textrm{NO}}\Big\},\\
                \st:=\Big\{\ket{U_\pm}_{\scriptscriptstyle\textrm{NO}},\,
                \ket{D_\pm}_{\scriptscriptstyle\textrm{NO}}\Big\}.
            \end{gathered}
        \end{equation}
        Focusing on $\so$, mappings $\kom{LR}\leftrightarrow\kop{LR}$ and $\ktm{LR}\leftrightarrow\ktp{LR}$ can be obtained by locally applying a $\pi$-PIPS to one of the two spatial modes, while a PIPS of $\pi/2$ achieves $\kom{NO}\leftrightarrow\kop{NO}$. Connections $\kom{LR}\leftrightarrow\ktm{LR}$ and $\kop{LR}\leftrightarrow\ktp{LR}$ can be obtained by means of a local PR performing the operation $\ket{\uparrow}\leftrightarrow\ket{\downarrow}$ on one mode. This set of local transformations relating Bell states were firstly introduced in Ref.\tilde\cite{Bennett1996}. We now extend them by noticing from Eq.\tilde\eqref{bsaction} that the non-locality generated by a BS can be employed to achieve the transformation $\kop{LR}\leftrightarrow\kom{NO}$.
        Considering $\st$, instead, mappings $\kum{NO}\leftrightarrow\kup{NO}$ and $\kdm{NO}\leftrightarrow\kdp{NO}$ can be realized by a local $\pi/2$-PIPS on one spatial mode, while $\kum{NO}\leftrightarrow\kdm{NO}$ and $\kup{NO}\leftrightarrow\kdp{NO}$ can be obtained by applying a PR to both modes. Sets $\so$ and $\st$ and the related intra-set PO relations are depicted in Fig\tilde\ref{poscheme}.

        We now show that a link between the two sets can be established by employing the (non PO) detector D described above. To move from $\so$ to $\st$, we start from state $\ktp{LR}\in\so$. We apply a PBS on one arbitrary spatial mode, placing D at the output of one of its ports before recombining the outputs in another PBS. Notice that such a Mach-Zehnder-like setup behaves as a \textit{polarization-sensitive}, non-absorbing, parity check detector. This allows to discriminate the component $\kluru$ of state $\ktp{LR}$ from the one $\kldrd$. Combining the two spatial modes in a BS now leads to either state $\kum{NO}\in\st$ or $\kdm{NO}\in\st$, respectively, as can be computed using Eq.\tilde\eqref{bsaction} and recalling that $\kluru=\ktp{LR}+\ktm{LR}$, $\kldrd=\ktp{LR}-\ktm{LR}$.
        To move from $\st$ to $\so$, instead, let us begin with $\kum{NO}\in\st$. Acting on it with a beam splitting operation, we obtain $(\ktp{LR}+\ktm{LR})/\sqrt{2}$. A PR set on the R spatial mode gives $(\kop{LR}+\kom{LR})/\sqrt{2}$, which is transformed by a second BS into $(\kom{NO}-\kom{LR})/\sqrt{2}$. The detector D can now be employed to discriminate the odd component ($\kom{LR}$) from the even one ($\kom{NO}$), both belonging to $\so$.
        Given the intra-set connections discussed above, we have thus found a link
        \begin{equation}
            \so\xleftrightarrow{PO + D}\st,
        \end{equation}
        which allows to transform any two arbitrary maximally entangled states in $\mathcal{B}$ one into the other.
        As these include the Bell singlet state, we thus conclude that the proposed scheme can be employed to prepare any maximally entangled state of two photonic qubits.

    \textit{Faulty parity check detector.}
        We conclude our analysis by accounting for possible errors occurring during the parity check detection.


        Errors may occur when the system state $\kom{LR}\bom{LR}$ is mistakenly detected as an even-parity state, and (or) when the system state $\rno$ in Eq.\tilde\eqref{rno} is wrongly detected as an odd-parity state. Accounting for these events amounts to substituting the previously defined projectors $\proj{LR}$ and $\proj{NO}$ with $\proj{LR}':=(1-\epsilon)\,\proj{LR}+\epsilon'\,\proj{NO}$ and $\proj{NO}':=(1-\epsilon')\,\proj{NO}+\epsilon\,\proj{LR}$, respectively, where error probabilities $\epsilon$, $\epsilon'$ are considered. Correspondingly, the system is projected into the states
        \begin{equation}
            \begin{aligned}
                \rlr'
                &=\frac{1}{4}\Big[(1-\epsilon)\,\kom{LR}\bom{LR}+3\,\epsilon'\,\rno\Big]/\plr',\\
                \rno'
                &=\frac{1}{4}\Big[3\,(1-\epsilon')\,\rno+\epsilon\,\kom{LR}\bom{LR}\Big]/\pno',
            \end{aligned}
        \end{equation}
        with related probabilities
        \begin{equation}
            \begin{aligned}
                \plr'&=(1-\epsilon)/4 + 3\epsilon'/4,\\
                \pno'&=3(1-\epsilon')/4 + \epsilon/4 =1-\plr'.
            \end{aligned}
        \end{equation}

\begin{figure}[t!]
            \includegraphics[width=0.8\columnwidth]{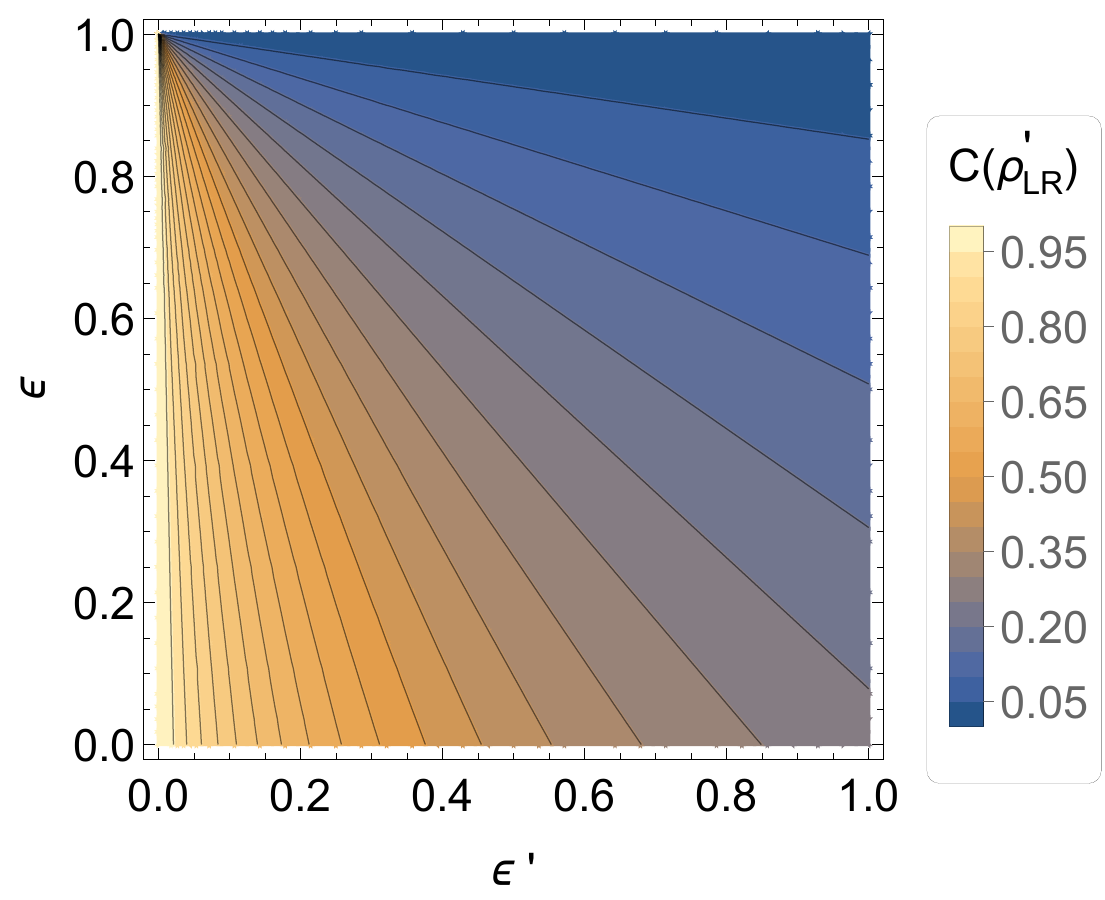}
            \centering
                \caption{Concurrence of the prepared state $\rlr'$, as a function of the error probabilities $\epsilon$ and $\epsilon'$ characterizing a faulty detection.}
        \label{concplot}
        \end{figure}

        
        It is now interesting to quantify the amount of quantum correlations present in the faulty state $\rlr'$ we collect. Since when no errors occur we expect to get the singlet state $\kom{LR}$, we focus on the entanglement in polarization. To do so, we use the concurrence\tilde\cite{concurrence}, ranging from $C(\rho)=0$ for separable states to $C(\rho)=1$ for maximally entangled ones, obtaining
        \begin{equation}
        \label{concurrence}
            C(\rlr')
            =\frac{1-\epsilon}{1-\epsilon+3\,\epsilon'}.
        \end{equation}
        Notice that the amount of entanglement in $\rlr'$ depends on both the error probabilities $\epsilon$ and $\epsilon'$, ranging from $C(\rlr')=0$ when $\epsilon=1$, to $C(\rlr')=1/(1+3\epsilon')$ when $\epsilon=0$. Fig.\tilde\ref{concplot} reports the concurrence $C(\rlr')$ as a function of $\epsilon$ and $\epsilon'$. 
        We remark that, however, not all the scenarios are relevant. When $\epsilon=1$, for example, it is enough to collect the photons when the detector signals an even parity state to achieve a state with nonzero entanglement (unless $\epsilon'=0$, too). 

    \textit{Conclusions.}
        In this work, we have introduced a procedure to robustly prepare maximally entangled states of two photonic qubits undergoing arbitrary local noise.
        
        The protocol employs PO transformations and a polarization-insensitive, non-absorbing, parity check detector to distil a Bell singlet state from a completely depolarized one. As the local depolarization of spatially distinguishable photons leads to the maximally mixed state regardless of the previous dynamics, the proposed scheme transforms any arbitrary initial state into the Bell singlet. In this way, the preparation is robust to the action of any local noise affecting the photons before their state is reset by the depolarization. We highlight that, in case a photon is lost during a noisy interaction, a new depolarized photon can be injected to recover the process.
        Employing a QND measurement, the protocol is iterative and prepares the desired state with a probability which scales exponentially with the number of repetitions, thus being asymptotically-deterministic.
        
        We have further introduced a formal equivalence based on PO transformations, showing that it allows to divide maximally entangled states of two qubits distributed over two distinct spatial modes in two sets of PO equivalent states. A connection between the two sets has been established employing the polarization-insensitive, non-absorbing, parity check detector previously discussed. As the Bell singlet state belongs to one of the two sets, we thus conclude that the proposed procedure allows for a robust generation of any arbitrary maximally entangled state of two photonic qubits.
        
        We emphasize that, to achieve the correct interference patterns, the PO transformations realized by BSs or PBSs require the two photons to be indistinguishable in all the degrees of freedom but the spatial one and, at most, the polarization. In light of this, the PO equivalence defined in this Letter can be ultimately interpreted as a connection between two synchronized sources of single photons satisfying the above requirement and the set of maximally entangled bipartite states.

        While an experimental implementation of the required PO transformations, including the realization of the depolarizing channels, can be efficiently implemented with commercially available devices, the realization of the polarization-insensitive, non-absorbing, parity check detector constitutes the main obstacle to be tackled. For this reason, we have analyzed the performances of our protocol when faulty detections are involved, studying the amount of quantum correlations between the two resulting photons as a function of the errors introduced by the detector. To this regard, we remark that the proposed scheme can still be employed substituting such a detector with commercially available single photon detectors performing a coincidence measurement on the two output modes, achieving the preparation of the desired maximally entangled state with probability $\plr=1/4$.

        We finally highlight that the reported results hold for any type of bosonic system, thus not being limited to photons. We foresee an extension of our procedure to fermions, clarifying the role of particle statistics in the preparation of entangled states. Moreover, we aim at widening the analysis of PO transformations to systems of $N>2$ particles, looking for a suitable generalization of the protocol presented in this work to prepare multipartite entangled states.

    \begin{acknowledgments}
            R.L.F. acknowledges support from European Union -- NextGenerationEU -- grant MUR D.M. 737/2021 -- research project ``IRISQ''.
            V.G. acknowledges financial support by MUR (Ministero dell'Universit\`{a} e della Ricerca) through the following projects: PNRR MUR project PE0000023-NQSTI, PRIN 2017 Taming complexity via Quantum Strategies: a Hybrid Integrated Photonic approach (QUSHIP) Id. 2017SRN-BRK, and project PRO3 Quantum Pathfinder.
    \end{acknowledgments}

    
%

\end{document}